\documentclass[12pt]{article}

\usepackage{amsmath}
\usepackage{amsfonts}
\usepackage{amssymb}
\usepackage{graphicx}
\usepackage{psfrag}
\usepackage{graphicx}
\usepackage{psfrag}

\newcommand{\be}{\begin{equation}}
\newcommand{\ee}{\end{equation}}
\newcommand{\ba}{\begin{eqnarray}}
\newcommand{\ea}{\end{eqnarray}}

\begin{document}
\begin{center}
{\bf EXACTLY SOLVABLE NON-SEPARABLE AND NON-DIAGONALIZABLE 2-DIM MODEL WITH QUADRATIC COMPLEX INTERACTION}\\
\vspace{1cm}
{\large \bf F. Cannata $^{1,}$\footnote{E-mail: cannata@bo.infn.it},
M. V. Iof\/fe $^{2,}$\footnote{E-mail: m.ioffe@pobox.spbu.ru},
D. N. Nishnianidze} $^{2,3,}$\footnote{E-mail: cutaisi@yahoo.com}\\
\vspace{0.5cm}
$^1$ Dipartimento di Fisica and INFN, Via Irnerio 46, 40126 Bologna, Italy.\\
$^2$ Saint-Petersburg State University,198504 Sankt-Petersburg, Russia\\
$^3$ Akaki Tsereteli State University, 4600 Kutaisi, Georgia
\end{center}
\vspace{0.5cm}
\hspace*{0.5in}
\vspace{1cm}
\hspace*{0.5in}
\begin{minipage}{5.0in}
{\small
We study a quantum model with non-isotropic two-dimensional oscillator potential but with additional quadratic interaction
$x_1x_2$ with imaginary coupling constant. It is shown, that for a specific connection between coupling constant
and oscillator frequences, the model {\it is not} amenable to a conventional separation of variables. The property
of shape invariance
allows to find analytically all eigenfunctions and the spectrum is found to be
equidistant. It is shown that the
Hamiltonian is non-diagonalizable, and the resolution of the identity must include also the corresponding associated functions.
These functions are constructed explicitly, and their properties are investigated. The problem of $R-$separation of
variables in two-dimensional systems is discussed.
}
\end{minipage}



\section*{\normalsize\bf 1. \quad Introduction.}
\vspace*{0.5cm}
\hspace*{3ex}
During last years, starting from a pioneering paper of C.Bender and S.Boettcher
\cite{bender0}, there is a growing interest to investigate  Quantum
Mechanics
with non-Hermitian Hamiltonians (see also \cite{moiseyev}) consistently. It was shown that under
definite assumptions the spectrum of
such Hamiltonians is real and a modified scalar product which provides unitary evolution can be built
for some models. For  comprehensive reviews, see papers \cite{bender-review}, \cite{mostafa-review}
and references therein.

With few exceptions \cite{bender2}, \cite{srilanka}, \cite{pseudo},
the analysis concerned one-dimensional Quantum Mechanics.
In particular,  most  results were obtained for a wide class of models, with
  unbroken $PT-$invariance \cite{bender0}, \cite{bender01}, \cite{bender02}, \cite{tateo}.
It can be considered as a modern generalization of conventional Quantum Mechanics to a non-Hermitian one.

In turn, the notion of the pseudo-Hermiticity:
\be
\eta H \eta^{-1} = H^{\dagger} \label{ps}
\ee
with $\eta$ a Hermitian invertible operator, allowed to define a more general class
of non-Hermitian systems with physically acceptable properties of energy spectra.
The most systematic investigation of pseudo-Hermiticity has been performed by
A.Mostafazadeh \cite{most} (see also \cite{ahmed}, \cite{japaridze}).
A suitable description of Hilbert space for such systems is given in terms of
biorthogonal basis, which consists of the eigenstates $|\Psi_n\rangle$ and $|\tilde\Psi_n\rangle$
of $H$ and $H^{\dagger},$ correspondingly.

It was found that some systems with complex potentials are naturally described by Hamiltonians
which are not diagonalizable. They correspond to the systems whose biorthogonal basis elements
do not provide {\it complete} basis in Hilbert space. In such a case, one has to
add the so-called associated functions to complete the basis, and Hamiltonian becomes block-diagonal with
some number of Jordan blocks of standard structure on its diagonal.

In Section 2 we formulate the two-dimensional model with complex potential having the form of second order polynomial
in $x_1,x_2$. Usually, such model is solved easily by means of linear transformation of coordinates
with subsequent separation of variables, maybe complex. This procedure was described, for example, in \cite{srilanka}.
But two peculiar cases of such polynomial potentials with special relation between constants
are beyond this scheme: they are not amenable to
separation of variables. Just such model is studied in Section 2. It is solved exactly: the whole energy
spectrum and corresponding wave functions are found analytically.
Instead of by separation of variables, which is
impossible here, the model is solved by means of shape invariance, a powerful
method introduced in the framework of SUSY Quantum Mechanics. In Section 3 we investigate
the properties of the constructed wave functions. We show that they do not realize a resolution of identity,
i.e. they
do not form complete basis, i.e. the Hamiltonian is not diagonalizable \cite{non-diag}.
The corresponding associated functions are also built analytically in this Section,
and their properties are studied in detail.
In Section 4, we discuss the conventional procedure of separation of variables in two-dimensional
Schr\"odinger equation both with real and with complex potentials. For the first case, the old results
of Eisenhart \cite{eisenhart} are reproduced, and for the second case, we prove that the model of previous Sections
does not allow the most general (nonlinear) algorithm of $R-$separation of variables \cite{miller}.

\section*{\normalsize\bf 2.\quad Two-dimensional complex oscillator.}
\vspace*{0.5cm}
\hspace*{3ex}
Let us consider the two-dimensional model with complex oscillator Hamiltonian:
\be
H=-\Delta^{(2)}+V(\vec x)=-\partial^2_1-\partial^2_2+\omega_1^2x_1^2+\omega_2^2x_2^2+2igx_1x_2.
\label{H}
\ee
Performing the linear complex transformation of variables $x_1,x_2$
\be
x_i=\sum_{j=1}^{2}a_{ij}y_j, \label{transf}
\ee
where $a_{ij}$ are complex elements of matrix $A,$ one may try to separate variables
in the Schrodinger equation:
\be
H\Psi(\vec x)=E\Psi(\vec x).
\label{schr}
\ee
It is necessary to obtain in  diagonal form both the Laplacian and the quadratic potential,
though in complex variables $y_i.$
As one can check, this is possible for generic values of parameters $\omega_i, g$ in (\ref{H}),  with
two exclusions.Indeed this is impossible, if the coupling constant is:
\begin{equation}\label{connection}
2g=\pm (\omega_1^2-\omega_2^2),
\end{equation}
when the Jacobian of (\ref{transf}) vanishes. Just this situation will be considered below in this paper,
and for definiteness we will choose the minus sign above.

We will use the complex variables $z=x_1+ix_2,\, \bar z=x_1-ix_2=z^{\ast},$ for which
\begin{equation}\label{Hz}
H=-4\partial_z\partial_{\bar z} +\lambda^2 z\bar z+g\bar z^2;\quad 2\lambda^2\equiv \omega_1^2+\omega_2^2>0;
\quad \lambda >0 .
\end{equation}
One can check that the Hamiltonian (\ref{Hz}) obeys the following property:
\begin{equation}\label{shape}
HA^{+}=A^{+}(H + 2\lambda); \quad HA^{-}=A^{-}(H - 2\lambda),
\end{equation}
with operators $A^{\pm}$ defined as:
\begin{equation}\label{A}
A^{\pm}\equiv \partial_z\mp \frac{\lambda}{2}\bar z.
\end{equation}
The equation (\ref{shape}) realizes the particular case of shape invariance \cite{shape},
the property appeared on the first
time in the framework of one-dimensional SUSY Quantum Mechanics \cite{SUSY} with real potentials. Shape invariance
of the Hamiltonian $H(x;a),$
which depends on a parameter $a$ means that this Hamiltonian satisfies the intertwining relations with some operators
$Q^{\pm}$:
\ba
H(x;a)Q^+&=&Q^+H(x;\tilde a)+R(a);  \quad \tilde a=\tilde a(a); \quad R(a)>0; \label{intertw1}\\
Q^-H(x;a)&=&H(x;\tilde a)Q^-+R(a).\label{intertw2}
\ea
This property provides a very elegant method
to solve the Schr\"odinger equation algebraically: practically, all known one-dimensional exactly-solvable
models are shape-invariant.
Recently this property was generalized to the two-dimensional Quantum Mechanics \cite{new}, where it gives usually a
quasi-exact-solvability of the model (i.e. analytical construction of a part of wave functions). For the case of
two-dimensional Morse potential with specific values of parameters, shape invariance helped to
find the whole spectrum and corresponding eigenfunctions \cite{exact}.

The Eq.(\ref{shape}) corresponds to the simplest variant of shape invariance: $\tilde a(a)=a$ and $R(a)=2\lambda.$
This case was investigated in one-dimensional Quantum Mechanics \cite{shape-our}, and
the same idea must work also in two-dimensional case.
Since the intertwined Hamiltonians in (\ref{intertw1}) coincide,
such Hamiltonians were called self-isospectral, and for one-dimensional case self-isospectrality
leads to an equidistant (oscillator-like) character of the spectrum.
In our present case of two-dimensional complex shape invariant potential,
the property (\ref{shape}) provides also the oscillator-like spectrum of $H:$
\begin{equation}\label{spectrum}
E_n=2\lambda (n+1).
\end{equation}
We deal here with unusual Quantum Mechanics - with complex potential and, even more, with non-diagonalizable
Hamiltonian (see the next Section). Nevertheless, the absence of singularities in (\ref{H})
ensures that nothing like well known "fall to the center" phenomena \cite{landau} is possible here (the formal proof
can be found in Appendix).
The normalizable bound state wave functions will be exponentially
decreasing at infinity, having no singularities. The corresponding spectrum is bounded from below, the ground state
with energy $E_0$ is denoted as $\Psi_{0,0}(\vec x)$ (it will be clear below, why we
use two indices for enumeration of $\Psi $). The excited levels correspond to $n=1,2,...$ in (\ref{spectrum}).

From the second
intertwining relation (\ref{shape}) it follows that $\Psi_{0,0}$ must be a zero mode of $A^-:$
\begin{equation}\label{zero}
A^-\Psi_{0,0}(z, \bar z)=0.
\end{equation}
Otherwise, the Hamiltonian $H$ would have the lower level $(E_0-2\lambda )$, contrary to our assumption on
$\Psi_{0,0}$ above. The solution
of zero mode equation (\ref{zero}) can be found in general form.
Indeed, it can be written in terms of a new function
$\tilde\Psi_{0}$ of $\bar z$ only:
$$
\Psi_{0,0}(z, \bar z)\equiv \exp{(-\frac{\lambda}{2}z\bar z)}\tilde\Psi_{0}(\bar z),
$$
and the Schr\"odinger equation (\ref{schr}) leads to the first order differential equation:
$$
2\lambda\bar z\tilde\Psi_{0}^{\prime}(\bar z)=(E_0-g\bar z^2-2\lambda)\tilde\Psi_{0}(\bar z).
$$
Its solution
$$
\tilde\Psi_0(\bar z)=(\bar z)^{(E_0-2\lambda )/2\lambda }\exp{[-(\frac{g}{4\lambda}\bar z^2)]}\nonumber
$$
has to be a single-valued function on a plane, i.e. to be $2\pi -$periodic in polar
angle $\varphi$ leading therefore to the spectrum (\ref{spectrum}).
Formally, infinitely many functions solve (\ref{zero}):
\begin{equation}\label{formal}
  \Psi_{n,0}(z, \bar z) =c_{n,0} \bar z^n\exp{(-az\bar z-b\bar z^2)};
  \quad n=0,1,2...,;\quad a\equiv\frac{\lambda}{2};\quad b\equiv\frac{g}{4\lambda},
\end{equation}
with normalization constants $c_{n,0}.$ The actual ground bound state corresponds
to $n=0$ in (\ref{formal}), and it has energy $2\lambda .$

All excited states can be built by the standard algebraic procedure of shape invariance:
\begin{equation}\label{excited}
\Psi_n(z, \bar z)=(A^+)^n \Psi_{0,0}(z, \bar z);\quad n=1,2,...
\end{equation}
with exactly the same energies as in (\ref{spectrum}). It is easy to check, that expressions in (\ref{formal})
and (\ref{excited}) coincide for all $n>0,$ i.e. $\Psi_n(z,\bar z)=\Psi_{n,0}(z,\bar z),$ being the zero
modes of operator $A^-.$
One can also prove that assuming the existence of any eigenstate
different from (\ref{excited}) we would obtain the level with real part lying below $E_0,$
in contradiction with the definition of the ground state
having the lowest real part of the energy.
Thus, the
whole spectrum of the system (\ref{H}) is exactly known - (\ref{spectrum}).
The corresponding wave functions $\Psi_{n,0}(z,\bar z)$ in (\ref{formal}) are known
analytically as well, but additional investigation
of their properties is required. It will be performed in the next Section.

\section*{\normalsize\bf 3.\quad Non-diagonalizability of the Hamiltonian.}

\subsection*{\normalsize\bf 3.1.\quad General scheme.}
\vspace*{0.5cm}
\hspace*{3ex}
The Quantum Mechanics with non-Hermitian Hamiltonians needs  a suitable modification of the scalar product
and resolution of identity to make the model self-consistent \cite{bender-review}, \cite{mostafa-review}.
In general, if the non-Hermitian
Hamiltonian $H$ commutes with some {\it antilinear}
operator $B,$ this may be used for definition of a new scalar product between arbitrary elements of Hilbert
space as follows:
\begin{equation}\label{B}
\langle\langle\Psi |\Phi\rangle\rangle \equiv \int (B\Psi)\Phi .
\end{equation}
In the ordinary Quantum Mechanics with real potentials just the customary complex conjugation plays the role
of such operator $B,$ and the Hamiltonian $H$ is Hermitian under such scalar product $\langle\Psi |\Phi\rangle .$

In the case of a general antilinear operator $B,$ the Hamiltonian $H$ obeys Hermiticity but with a scalar
product (\ref{B}). Symbolically:
\begin{equation}\label{herm}
\langle\langle\Psi|H|\Phi\rangle\rangle = \int (B\Psi)H\Phi = \int (HB\Psi)\Phi =
\int (BH\Psi)\Phi = \langle\langle H\Psi|\Phi\rangle\rangle ,
\end{equation}
where double integration by parts and vanishing of off-integral terms were used to move $H$ to the left under
the integral. In the two-dimensional case, the Ostrogradsly-Gauss theorem (a two-dimensional analog of two integrations by parts)
allows to move $H$ as well, due to exponential decreasing of all wave functions on large contour. Many one-dimensional
models
with non-Hermitian Hamiltonians obey the so-called $PT-$symmetry \cite{bender0}, \cite{bender-review}, cite{bender01},
\cite{bender02},  where the role of $B$ is served by the antilinear
symmetry operator $PT$ of simultaneous time and coordinate reflections. In more general situation, non-Hermitian
Hamiltonians may be pseudo-Hermitian \cite{mostafa-review}
(the definition in operatorial form was given in (\ref{ps})). Then the operator $B$
can be chosen as $B\equiv \eta T,$ and the scalar product (\ref{B}) coincides with $\eta -$scalar-product commonly
used in the literature. The pseudo-Hermitian Hamiltonians (\ref{ps}) are Hermitian being considered
under this scalar product.

In our case of two-dimensional coordinate space, such antilinear operator, keeping $H$ in (\ref{H}) invariant, can
be chosen either as $P_1T$ or as $P_2T,$ where $P_1$ means $x_1\leftrightarrow -x_1,$ and $P_2$ means
$x_2\leftrightarrow -x_2.$ Let us choose the second option for definiteness. The wave functions $\Psi_{n,0}(z,\bar z)$ are
simultaneously the eigenfunctions of $P_2T$ with unique eigenvalue $+1.$ For such choice of the operator $B,$
the scalar product $\langle\langle\Psi |\Phi\rangle\rangle$ becomes an integral over the
product $\int\Psi \Phi ,$ instead of the $\int\Psi^{\star}\Phi$ in the ordinary Quantum Mechanics.

Now the important property of the wave functions under
a new scalar product has to be investigated - are the corresponding norms positively
definite or not. The norms of the basic states $\Psi_{n,0}$ can be calculated explicitly:
\ba
&&\langle\langle\Psi_{n,0} |\Psi_{n,0}\rangle\rangle = \int (\Psi_n)^2d^2x =
c^2_{n,0}\int \bar z^{2n}\exp{[-2(az\bar z + b\bar z^2)]}dzd\bar z = \nonumber\\
&&=c^2_{n,0}\int \bar z^{2n}\exp{[-2(az\bar z + b\bar z^2)]}dzd\bar z=
c^2_{n,0}(-\frac{1}{2}\partial_b)^n\int\exp{[-2(az\bar z + b\bar z^2)]}dzd\bar z =\nonumber\\
&&=c^2_{n,0} (-\frac{1}{2} \partial_b)^n(\frac{\pi}{2a})=\frac{\pi c^2_{n,0}}{2a}\delta_{n0}.
\label{norm}
\ea
Thus, only the ground bound state $\Psi_{0,0}$ is normalizable state with a positive norm. All excited wave functions have
zero norms (are "self-orthogonal"), prohibiting the usual resolution of identity in terms of complete
set of eigenfunctions of $H.$ Such
situation was investigated in one-dimensional Quantum Mechanics with complex potentials
in a series of papers \cite{non-diag}:
the zero norm of wave function signals that one deals with a non-diagonalizable Hamiltonian. Then it is necessary
to build the so-called associated functions which participate in a resolution of identity and complete the basis.

The adequate formalism for diagonalizable non-Hermitian Hamiltonians is the so-called
biorthogonal basis in the Hilbert space
\cite{mostafa-review}, \cite{most}.
This basis includes two types of states $|\Psi_n\rangle,\,|\tilde\Psi_n\rangle,$ such that:
\begin{equation}\label{basis}
H|\Psi_n\rangle = E_n |\Psi_n\rangle;\,\, H^{\dag}|\tilde\Psi_n\rangle =E^{\star}|\tilde\Psi_n\rangle;\,\,
\langle\tilde\Psi_n|\Psi_m\rangle = \langle\Psi_m|\tilde\Psi_n\rangle = \delta_{nm}
\end{equation}
with decompositions:
\begin{equation}\label{decomp}
I=\Sigma |\Psi_n\rangle\langle\tilde\Psi_n|;\quad H=\Sigma E_n|\Psi_n\rangle\langle\tilde\Psi_n|.
\end{equation}
In coordinate representation, one can take $\tilde\Psi_n(\vec x)=\Psi_n^{\star}(\vec x),$ so that the scalar product is:
\begin{equation}\label{prod}
  \langle\tilde\Psi_n|\Psi_m\rangle = \int \Psi_n(\vec x)\Psi_m(\vec x)d^2x =
  \langle\langle\Psi_n|\Psi_m\rangle\rangle =\delta_{nm}.
\end{equation}
This formalism has been discussed in \cite{mostafa-review}, \cite{most}, \cite{non-diag}, for a particular explicit
calculations see also \cite{levai} (in one-dimensional case) and \cite{pseudo} (in two dimensions).

The formalism becomes more complicated \cite{non-diag} in the case of quantum systems
with non-diagonalizable non-Hermitian Hamiltonians, like our (\ref{H}).
For such systems, the basis (\ref{basis}) is not complete, and the decompositions
(\ref{decomp}) do not work. In order to improve the situation, every self-orthogonal
wave function $\Psi_{n,0},\, n\geq 1$ with zero norm must be accompanied with a set of $p_n-1$
associated functions $\Psi_{n,k},\, k=1,2,...,p_n-1.$ It must be clear now, why
notations with two indices of wave functions were introduced above. By definition, these functions obey:
\begin{equation}\label{assoc}
 (H-E_n)\Psi_{n,k}=\Psi_{n,k-1};\quad k=1,2,...,p_n-1,
\end{equation}
where all functions are supposed to be normalizable, and
the operator H, when applied to these functions, preserves this property.
Here we restrict ourselves for simplicity with the case when
each self-orthogonal eigenfunction $\Psi_{n,0},\, n=1,2,...$
is accompanied by {\it only one} set of associated functions $\Psi_{n,k},\, k=1,2,...,p_n-1.$

Similarly to the scheme of previous paragraph, the partner eigenfunctions
$\tilde\Psi_{n,0}$ also are accompanied by their associated functions $\tilde\Psi_{n,k},\, k=1,2,...,p_n-1.$
Practically, it is convenient to numerate the functions $\tilde\Psi ,$
identifying them with $\Psi^{\star},$ as follows:
\begin{equation}\label{renumber}
\tilde\Psi_{n,p_n-k-1}=\Psi^{\star}_{n,k}\quad k=0,1,2,...p_n-1.
\end{equation}
With these notations, according to the general formalism which was illustrated in detail for some one-dimensional
models \cite{non-diag}, the scalar product (\ref{prod}) in the extended biorthogonal basis must be:
\ba
&&\langle\langle\Psi_{n,k}|\Psi_{m,l}\rangle\rangle = \langle\tilde\Psi_{n,k}|\Psi_{m,l}\rangle =
\int \Psi_{n,k}(\vec x)\Psi_{m,l}(\vec x)d^2x =\delta_{nm}\delta_{k\, (p_n-l-1)};\label{product}\\
&&k=0,1,...,p_n-1;\,\, l=0,1,...,p_m-1 .\nonumber
\ea
Correspondingly, the generalized decompositions must be:
\ba
I&=&\sum_{n=0}^{\infty}\sum_{k=0}^{p_n-1} |\Psi_{n,k}\rangle\rangle\langle\langle\Psi_{n,p_n-k-1}|;\label{decomp1} \\
H&=&\sum_{n=0}^{\infty}\sum_{k=0}^{p_n-1} E_n|\Psi_{n,k}\rangle\rangle\langle\langle\Psi_{n,p_n-k-1}|
+ \sum_{n=0}^{\infty}\sum_{k=0}^{p_n-2} |\Psi_{n,k}\rangle\rangle\langle\langle\Psi_{n,p_n-k-2}|.\label{decomp2}
\ea
The Hamiltonian $H$ is clearly non-diagonal, but block-diagonal. Each block - Jordan cell of standard form
(see (\ref{decomp2})) - has dimensionality $p_n.$

\subsection*{\normalsize\bf 3.2.\quad The specific model: Non-Hermitian two-dimensional oscillator.}
\vspace*{0.5cm}
\hspace*{3ex}
The eigenfunctions $\Psi_{n,0}$ were found analytically in the previous Section (see Eq.(\ref{formal})). All these
functions for $n\geq 1$ were shown to be self-orthogonal, and therefore, some associated functions
must be properly taken into account.
In this Subsection we are going to investigate the properties of these functions, and in particular, to check
the relations (\ref{product}) and to find the values of $p_n.$

First of all, we will prove that scalar products (\ref{product}) vanish for different energy levels $E_n,\,E_m,$ i.e. that
$\langle\langle\Psi_{n,k}|\Psi_{m,l}\rangle\rangle =0$ for $n\neq m.$ The proof is by induction
in $k,l$ with essential use of
pseudo-Hermiticity of $H.$ Indeed, as in ordinary Quantum Mechanics:
$$
0=\langle\langle\Psi_{n,0}|H|\Psi_{m,0}\rangle\rangle-\langle\langle H\Psi_{n,0}|\Psi_{m,0}\rangle\rangle
=(E_m-E_n)\langle\langle\Psi_{n,0}|\Psi_{m,0}\rangle\rangle ,
$$
i.e. wave functions with different energies are orthogonal. Analogously, because of definition (\ref{assoc}):
\ba
0&=&\langle\langle\Psi_{n,0}|(H-E_m)|\Psi_{m,1}\rangle\rangle -
\langle\langle (H-E_m)\Psi_{n,0}|\Psi_{m,1}\rangle\rangle=\nonumber\\&=&
\langle\langle \Psi_{n,0}|\Psi_{m,0}\rangle\rangle -(E_n-E_m)\langle\langle\Psi_{n,0}|\Psi_{m,1}\rangle\rangle \nonumber ,
\ea
and the scalar products between wave functions and first associated functions for different $E_n,\,E_m$ also vanish:
$$
\langle\langle \Psi_{n,0}|\Psi_{m,1}\rangle\rangle = \langle\langle\Psi_{n,1}|\Psi_{m,0}\rangle\rangle =0.
$$
The procedure can be continued further leading to orthogonality of all functions with different $n,m.$

Now we have to consider the wave functions and associated functions, which belong to the same value $n.$
For the first associated function $\Psi_{n,1}$, the defining equation (\ref{assoc})
can be solved explicitly in a general form with two arbitrary constants:
\ba
\Psi_{n,1}(z,\bar z)=&&\biggl[a_{n,1}z\bar z^{n-1}-a_{n,1}\frac{n-1}{\lambda}\bar z^{n-2}+c_{n,1}\bar z^n
+\frac{1}{2\lambda}(c_{n,0}-\frac{2a_{n,1}g}{\lambda})\bar z^n\ln{\bar z}\biggr]\cdot\nonumber\\
&&\cdot\exp{[-(\frac{\lambda}{2}z\bar z+\frac{g}{4\lambda}\bar z^2)]} . \nonumber
\ea
One of integration constants is defined by physical requirement for wave functions to be single-valued in the plane:
\begin{equation}\label{ln}
a_{n,1}=\frac{\lambda c_{n,0}}{2g}=\frac{c_{n,0}}{8b},
\end{equation}
leading to:
\begin{equation}\label{ffirst}
\Psi_{n,1}(z,\bar z)=[\frac{c_{n,0}}{8b}z\bar z^{n-1}-\frac{(n-1)c_{n,0}}{16ab}\bar z^{n-2}+c_{n,1}\bar z^n]
\exp{[-(az\bar z+b\bar z^2)]}.
\end{equation}
The second integration constant $c_{n,1}$ reflects the obvious fact (see (\ref{assoc})), that $\Psi_{n,1}$
is defined up to a solution of homogeneous Schr\"odinger equation $c_{n,1}\Psi_{n,0}.$ This additional term
must be defined by a suitable "normalization", i.e. by conditions (\ref{product}). One can check, that
the normalization conditions (\ref{product}) fix uniquely not only $c_{n,1},$ but all
higher additional terms as well.

To find the norm of $\Psi_{n,1}$ and some other scalar products below, we shall calculate a class of
two-dimensional integrals of the form:
\begin{equation}\label{int}
I_{N,M}\equiv\int z^N\bar z^M \exp{[-2(az\bar z+b\bar z^2)]}dzd\bar z
\end{equation}
with positive constants $a,b$ and integer $N,M.$ These integrals vanish for odd values of $(N+M)$
due to antisymmetry under a
space reflection $(x_1,x_2) \rightarrow -(x_1,x_2),$ i.e.  $I_{N,M}=0$ for $(N+M)=2s+1.$ For even values of $(N+M),$
we start from the basic integral \cite{prudnikov}:
\begin{equation}\label{basic}
I(a,b,c)=\int \exp{[-2(az\bar z+b\bar z^2+cz^2)]}dzd\bar z=\pi\delta^{-1};\,\,\delta\equiv 2\sqrt{(a^2-4bc)}.
\end{equation}
Then, the required $I_{N,M}$ can be calculated by suitable differentiations of $I(a,b,c)=I(\delta):$
\ba
&&I_{2n,2(n+k)}=\int (z\bar z)^{2n}\bar z^{2k} \exp{[-2(az\bar z+b\bar z^2+cz^2)]}dzd\bar z_{|c=0}=\nonumber\\
&&=[(-\frac{1}{2}\partial_a)^{2n}(-\frac{1}{2}\partial_b)^{k}I(\delta )]_{|c=0}=0;\,\,k>0 \label{I-1}\\
&&I_{2(n+k),2n}=\int (z\bar z)^{2n}z^{2k} \exp{[-2(az\bar z+b\bar z^2+cz^2)]}dzd\bar z_{|c=0}=\nonumber\\
&&[(-\frac{1}{2}\partial_a)^{2n}(-\frac{1}{2}\partial_c)^{k}I(\delta )]_{|c=0}=
\pi (-1)^k2^{-(2n+1)}(2k+1)!(2k+1)_{2n}b^ka^{-(2k+2n+1)} ; \label{I-2}\\
&&I_{2n+1,2(n+k)+1}=\int (z\bar z)^{2n+1}\bar z^{2k} \exp{[-2(az\bar z+b\bar z^2+cz^2)]}dzd\bar z_{|c=0}=\nonumber\\
&&=[(-\frac{1}{2}\partial_a)^{2n+1}(-\frac{1}{2}\partial_b)^{k}I(\delta )]_{|c=0}=0; \,\, k>0 \label{I-3}\\
&&I_{2(n+k)+1,2n+1}=\int (z\bar z)^{2n+1} z^{2k} \exp{[-2(az\bar z+b\bar z^2+cz^2)]}dzd\bar z_{|c=0}=\nonumber\\
&&=[(-\frac{1}{2}\partial_a)^{2n+1}(-\frac{1}{2}\partial_c)^{k}I(\delta )]_{|c=0}=
\pi (-1)^k2^{-(2n+2)}(2k+1)!(2k+1)_{2n+1}b^ka^{-(2k+2n+2)}, \label{I-4}
\ea
In particular, it is clear that $I_{N,M}=0$ for $M>N.$ The listed integrals allow to check that
for the first excited level $n=1$ functions
$\Psi_{1,0},\,\Psi_{1,1},$ in addition to self-orthogonality of $\Psi_{1,0},$
satisfy (up to normalization factors) the relations (\ref{product}) with the value $p_1=2:$
$$
\int\Psi_{1,1}\Psi_{1,0}dzd\bar z=1;\quad\int\Psi_{1,1}\Psi_{1,1}dzd\bar z=0,
$$
where one has to take $c_{1,1}=c_{1,0}/8a$ in (\ref{ffirst}) and $c^2_{1,0}=32a^2b/\pi $
in (\ref{formal}).
For  the general value $n\geq 2,$ the norm of the first associated functions
$\int (\Psi_{n,1})^2dzd\bar z=Const\cdot\delta_{n,2}.$ Comparing with the orthogonality relations (\ref{product}),
this signals that $p_2=3,$ and vice versa, $ p_n\neq 3$ for $n > 2$.

Calculation of scalar products for higher order associated functions $\Psi_{n,k}$ with $k\geq 2$ will be performed
by an alternative method, where the main role is played by the last associated functions $\Psi_{n,p_n-1}.$
From the definition (\ref{assoc}), the following expression for $\Psi_{n,k}$ in terms
of $\Psi_{n,p_n-1}$ can be easily derived:
\begin{equation}\label{ass}
  \Psi_{n,k}=(H-E_n)^{p_n-k-1}\Psi_{n,p_n-1}; \,\,k=0,1,...,p_n-1.
\end{equation}
The dimension $p_n-1$ of
Jordan cell and the highest associated function $\Psi_{n,p_n-1}$ must be derived by solving the equation:
\begin{equation}\label{last}
   (H-E_n)^{p_n}\Psi_{n,p_n-1}=0,
\end{equation}
by subsequent calculation of $\Psi_{n,k}$ along (\ref{ass}), and finally, by
checking the required scalar products (\ref{product}).

We shall prove now that $p_n=n+1$ and the following solution of Eq.(\ref{last})
 \be
 \Psi_{n,n}=c_n\Omega_n\exp{[-(az\bar z+b\bar z^2)]}, \quad
 \Omega_n\equiv(az+b\bar{z})^n,\quad n>0,\label{22}
 \ee
satisfy all necessary conditions above (constants of normalizations $c_n$ will be defined below).

In order to prove these statements, one has to check straightforwardly the following relations:
\ba
 (H-E_m)\cdot\Psi_{0,0}=\Psi_{0,0}\cdot 4\biggl(-\partial_{z}\partial_{\bar{z}}+
 a(z\partial_{z}+\bar{z}\partial_{\bar{z}}-m)+2b\bar{z}\partial_z\biggr)\equiv \Psi_{0,0}\cdot D_m, \label{44}
\ea
\ba
 (D_m\Omega_n)=4a\biggl(-bn(n-1)\Omega_{n-2}+
 (n-m)\Omega_n+2bn\bar{z}\Omega_{n-1}\biggr),\label{55}
 \ea
 \ba
 D_m\cdot\bar{z}^j=\bar{z}^jD_m-4j\bar{z}^{j-1}\partial_z+4aj\bar{z}^j.\label{66}
 \ea
By the method of mathematical induction, it can be derived from (\ref{44})-(\ref{66}), that:
$$
 D_n^k\Omega_n=(2ab)^k\sum_{i=0}^k\biggl(n-(2k-i-1)\biggr)_{2k-i}\alpha_i^{(k)}
 \bar{z}^i\Omega_{n-(2k-i)},
$$
where:
$$
 (a)_k=a(a+1)...(a+k-1);\,\, \alpha_0^{(k)}=(-2)^k,\quad \alpha_k^{(k)}=2^{2k},\,\,
 \alpha_{0<i<k}^{(k)}=\frac{(-2)^i(k-(i-1))_i \alpha_0^{(k)}}{i!},
$$
 and the coefficients $\alpha_j^{(k)}$ satisfy the system of equations:
 \ba
 2(i-k)\alpha_i^{(k)}=(i+1)\alpha_{i+1}^{(k)},\quad 0\leq i<k;\nonumber\\
 \alpha_{k+1}^{(k+1)}=4\alpha_k^{(k)},\quad \alpha_0^{(k+1)}=-2\alpha_0^{(k)};\nonumber\\
 2(2\alpha_{i-1}^{(k)}-\alpha_i^{(k)})=\alpha_i^{(k+1)},\quad 1\leq i\leq k.\nonumber
 \ea
Thus,
 \be
 (H-E_n)^k\Psi_{n,n}=c_n(2ab)^k\exp{[-(az\bar z+b\bar z^2)]}\sum_{i=0}^k\alpha_i^{(k)}\biggl(n-(2k-i-1)\biggr)_{2k-i}
 \bar{z}^i\Omega_{n-(2k-i)},\label{1010}
 \ee
 and for $k=n,$ only one coefficient of $\bar{z}^i$ does not vanish: $i=k=n.$
 Therefore,
 \be
 (H-E_n)^n\Psi_{n,n}=c_n(8ab)^nn!\bar{z}^n\exp{[-(az\bar z+b\bar z^2)]}=\frac{c_n(8ab)^n n!}{c_{n,0}}\Psi_{n,0}.\label{11}
 \ee
 The action of one more operator $(H-E_n)$ onto this equation leads just to
 required Eq.(\ref{last}). The
norm of $\Psi_{n,n}$ is calculated using the explicit expression (\ref{22}) and the integrals
(\ref{I-1}) - (\ref{I-4}):
\ba
&&\langle\langle\Psi_{n,n}|\Psi_{n,n}\rangle\rangle =
\int dzd\bar z(az+b\bar z)^{2n}\exp{[-2(az\bar z+b\bar z^2+cz^2)]}_{c=0}=\nonumber\\
&&=\int dzd\bar z(-\frac{1}{2})^n(a^2\partial_c+2ab\partial_a+b^2\partial_b)^n
\exp{[-2(az\bar z+b\bar z^2+cz^2)]}_{c=0}=\nonumber\\
&&=(-\frac{1}{2})^n[(a^2\partial_c+2ab\partial_a+b^2\partial_b)^nI(\delta)]_{c=0}=0. \nonumber
\ea
The action of the operator $(a\partial_a+b\partial_b)$ onto the vanishing integral above leads to the useful expressions
for arbitrary integer $n,k:$
\begin{equation}\label{nnn}
  \int dzd\bar z (az+b\bar z)^n\bar z^k\exp{[-2(az\bar z+b\bar z^2)]}=\delta_{nk}\frac{\pi n!}{2^{n+1}a}.
\end{equation}

Equations (\ref{1010}) for $k<n,$ together with Eqs.(\ref{ass}) and (\ref{nnn}), allow to analyze
other scalar products of associated functions with the same energy. Indeed, since
according to (\ref{1010}) $\Psi_{n,n-k}$ is a linear combination ($\beta_i-$ are combinations of constants
entering (\ref{1010})):
$$
 \Psi_{n,n-k}=\sum_{i=0}^k\beta_i\bar{z}^i\Omega_{n-2k+i}\Psi_{0,0},
$$
the required scalar products are:
\ba
 &&\langle\langle\Psi_{n,n-k},\Psi_{n,n-k'}\rangle\rangle
 =\int dzd\bar z\sum_{i=0}^k\sum_{j=0}^{k'}\beta_i\beta'_j\bar{z}^{i+j}
 \Omega_{n-2k+i}\Omega_{n-2k'+j}\exp{[-2(az\bar z +b\bar z^2)]}=\nonumber\\
 &&=\int dzd\bar z\sum_{i=0}^k\sum_{j=0}^{k'}
 \beta_i\beta'_j\bar{z}^{i+j}(az+b\bar{z})^{2n-2(k+k')+i+j}
 \exp{[-2(az\bar{z}+b\bar{z}^2)]}.\nonumber
 \ea
Therefore, due to Eq.(\ref{nnn}), the following choice of $c_{n,0}$ in (\ref{formal}) and $c_n$ in
(\ref{22}):
$$
c_n=\frac{c_{n,0}}{(8ab)^nn!},\quad c_{n,0}^2=\frac{2a(16ab)^n}{\pi}
$$
provides normalized scalar products
\be
 \langle\langle\Psi_{n,k},\Psi_{n,k'}\rangle\rangle=\delta_{k,n-k'}, \label{7777}
 \ee
as it should be for $p_n=n+1.$
Summarizing this Subsection, we proved that the dimension of Jordan cell corresponding to
energy $E_n$ depends on $n,$ namely, $p_n=n+1,$ and
all constructed associated functions, after suitable normalization,
provide the necessary scalar products (\ref{product}).

\section*{\normalsize\bf 4.\quad Non-separability of variables.}
\vspace*{0.5cm}
\hspace*{3ex}
It was already noted in Section 2, that the Schr\"odinger equation with (complex) potential of
second order in $x_1,\,x_2,$ just in the case (\ref{connection}), does not allow an ordinary separation
of variables by means of linear transformation of coordinates. A more general question will be studied
in this Section: whether any {\it nonlinear} transformation can provide the separation. Actually, we are interested
in opportunity to perform the so-called $R-$separation of variables \cite{miller}, which is explored, for example,
in three-dimensional problems
with central forces. In two-dimensional context, $R-$ separation means that coordinates $x_1,\, x_2$
can be mapped (and the mapping is invertible) to new (complex, in general) variables:
\begin{equation}\label{xq}
q_1=Q_1(x_1, x_2), \,\, q_2=Q_2(x_1, x_2);\quad x_1=X_1(q_1, q_2),\,\, x_2=X_2(q_1, q_2),
\end{equation}
so that the Hamiltonian takes the form:
\begin{equation}\label{R}
H=\frac{1}{\tau_1{(q_1)}+\tau_2{(q_2)}}\biggl[-(\partial_{q_1}^2+\partial_{q_2}^2)+\mu_1(q_1)\partial_{q_1}+
\mu_2(q_2)\partial_{q_2}+\vartheta_1(q_1)+\vartheta_2(q_2)\biggr],
\end{equation}
with arbitrary functions $\tau_i,\, \mu_i,\, \vartheta_i.$ In such a case, the problem splits onto two
one-dimensional problems. The separation of variables in two- and three-dimensional Schr\"odinger equation
with real potentials was investigated by L.P.Eisenhart \cite{eisenhart}, where the exhaustive lists
of corresponding coordinate systems $q_i(\vec x)$ and potentials $V(\vec x)$ were found:
eleven systems in three dimensions and three systems in two-dimensional case. Because of complexity
of potential, we are interested here in generalization of these results on
two-dimensional $R-$separation: both potentials $V(\vec x)$ and the new coordinates $q_1, q_2$ may be complex.

The change of variables (\ref{xq}) in the kinetic part of
$H=-(\partial_{x_1}^2+\partial_{x_2}^2)+V(x_1, x_2)$ gives the following conditions:
\ba
&&(\partial_{x_1}Q_1) (\partial_{x_1}Q_2) + (\partial_{x_2}Q_1) (\partial_{x_2}Q_2)=0 \label{xx}\\
&&(\partial_{x_1}Q_1)^2 + (\partial_{x_2}Q_1)^2 = (\partial_{x_1}Q_2)^2 + (\partial_{x_2}Q_2)^2
=-\frac{1}{\tau_1(q_1)+\tau_2(q_2)} \label{xxx},
\ea
which lead, in particular, to relation:
\begin{equation}\label{gauge}
\partial_{x_1}Q_2=\partial_{x_2}Q_1.
\end{equation}
Actually, the opposite sign in the r.h.s. of (\ref{gauge})
is also possible, but the final results will be the same.
The general solution of (\ref{gauge}) is expressed in terms of an arbitrary complex function $G$:
\begin{equation}\label{G}
  Q_1=\partial_{x_2}G(x_1, x_2);\quad Q_2=\partial_{x_1}G(x_1, x_2).
\end{equation}
After substitution of (\ref{G}) back into (\ref{xx}),
\begin{equation}\label{GG}
  (\partial_{x_1}\partial_{x_2}G(x_1, x_2))\cdot (\partial_{x_1}^2+\partial_{x_2}^2)G(x_1, x_2)=0,
\end{equation}
we have two options:
\ba
&&\partial_{x_1}\partial_{x_2}G(x_1, x_2)=0; \label{option1}\\
&&(\partial_{x_1}^2+\partial_{x_2}^2)G(x_1, x_2)=0. \label{option2}
\ea

For the first option,the variable $q_1$ depends on $x_1$ only, and analogously, $q_2$ on $x_2.$ Then from
(\ref{xxx}),
$$
\biggl(\partial_{x_1}Q_1(x_1)\biggr)^2_{|x_1=X_1(q_1)}=\biggl(\partial_{x_2}Q_2(x_2)\biggr)^2_{|x_2=X_2(q_2)}=
-\frac{1}{\tau_1(q_1)+\tau_2(q_2)},
$$
for which only non-interesting solutions exist: $\tau_i(q_i)=Const,$ and correspondingly, $Q_i(x_i)$ are
linear complex functions of $x_i.$

The second option (\ref{option2}) gives more interesting general solution:
\ba
&&G(x_1, x_2)=m(z)+n(\bar z); \quad z=x_1+ix_2;\,\, \bar z=x_1-ix_2\nonumber\\
&&q_1=Q_1(x_1, x_2)=m^{\prime}(z)+n^{\prime}(\bar z);\quad
q_2=Q_2(x_1, x_2)=i(m^{\prime}(z)-n^{\prime}(\bar z));\nonumber\\
&&z=Z(q_-);\quad \bar z=\tilde Z(q_+)=\biggl(Z(q_-)\biggr)^{\star};
\quad q_+\equiv q_1+iq_2;\,\, q_-\equiv q_1-iq_2 \nonumber
\ea
in terms of new functions $m, n, Z, \tilde Z.$ In order to investigate, whether variables $q_1, q_2$ allow
separation of variables in the Schr\"odinger equation, we substitute this solution in the
kinetic part of Hamiltonian:
$$
(\partial^2_{x_1}+\partial^2_{x_2})=
4m(z)n(\bar z)(\partial^2_{q_1}+\partial^2_{q_2})_{|z=Z(q_-); \bar z=\tilde Z(q_+)}.
$$
The separation of variables (\ref{R}) in Laplacian is possible only if
$$
\bigl(4m(z)n(\bar z)_{|z=Z(q_-); \bar z=\tilde Z(q_+)}\bigr)^{-1} = \tau_1{(q_1)}+\tau_2{(q_2)},
$$
i.e., if:
$$
\partial_{q_1}\partial_{q_2}\bigl(m^{-1}(Z(q_-))n^{-1}(\tilde Z(q_+)\bigr) = 0.
$$
Thus, one obtains two ordinary differential equations:
\begin{equation}\label{gamma}
\frac{\bigl(n^{-1}(q_+)\bigr)^{\prime\prime}}{n^{-1}(q_+)} =
\frac{\bigl(m^{-1}(q_-)\bigr)^{\prime\prime}}{m^{-1}(q_-)} = \gamma^2;
\quad \gamma = const.
\end{equation}

After straightforward calculations, the case $\gamma =0$ leads to the following solutions:
\be
z=\alpha_1 q_-^2 + \beta_1 q_- + \kappa_1;\quad \bar z=\alpha_2 q_+^2 + \beta_2 q_+ + \kappa_2 , \label{alpha}
\ee
where $\alpha_i, \beta_i, \kappa_i $ are constants. Here $z=x_1+ix_2, \bar z=x_1-ix_2$ are still mutually conjugate,
although $q_+,\, q_-$ are, in general, not necessarily conjugate. If we are interested, similarly to \cite{eisenhart},
only in real variables $q_i,$ and therefore $q_+=q_-^{\star},$ then the constants in (\ref{alpha}) are mutually conjugate:
$\alpha_1=\bar\alpha_2\equiv \alpha ,\, \beta_1=\bar\beta_2\equiv\beta ,\, \kappa_1=\bar\kappa_2\equiv\kappa .$ In this case
for $\alpha\neq 0$ by means of a suitable constant shifts
of $q_i$ and $x_i,$ without loss of generality, one can made $\beta =\kappa =0.$
Choosing also the special scale, namely, $\alpha =1/2,$ we obtain:
\begin{equation}\label{qq}
q^2_{1,2}=\pm x_1+(x_1^2+x_2^2)^{1/2};\quad x_1=\frac{q_1^2-q_2^2}{2};\,\, x_2=q_1q_2,
\end{equation}
and
\begin{equation}\label{ttt}
\tau (q_i)=q_i^2;\quad \tau_1(q_1)+\tau_2(q_2)=2(x_1^2+x_2^2)^{1/2}.
\end{equation}
The potential takes the form:
\begin{equation}\label{S3}
V(x_1, x_2)=(x_1^2+x_2^2)^{-1/2}[f(x_1+(x_1^2+x_2^2)^{1/2})+g(-x_1+(x_1^2+x_2^2)^{1/2})].
\end{equation}
Expressions (\ref{qq}) - (\ref{S3}) coincide exactly with the case III of Eisenhart \cite{eisenhart}.

For $\alpha_1=\alpha_2 =0$ in (\ref{alpha}), the transformation $\vec x\rightarrow \vec q$ describes
linear transformations to complex coordinates $q_1, q_2,$ which were mentioned in Section 2. For complex
$q_1, q_2$ the general form $(\ref{alpha})$ is allowed. For example, $\alpha_1 =0,\, \alpha_2\neq 0$ among others.

The case $\gamma \neq 0$ gives from (\ref{gamma}):
\begin{equation}\label{sigma}
z=\sigma_1\exp{(\gamma q_-)} - \delta_1\exp{(-\gamma q_-)};\quad
\bar z =\sigma_2\exp{(\gamma q_-)} - \delta_2\exp{(-\gamma q_-)}.
\end{equation}
For the case of real $q_1, q_2$, the constant $\gamma $ is real,
and $\sigma_1=\bar\sigma_2\equiv\sigma,\,\, \delta_1=\bar\delta_2\equiv\delta .$ Two options for (\ref{sigma})
must be considered separately. If $\delta =0,$ by means of rotations in $(x_1, x_2)-$plane one obtains:
\ba
&&x_1=\sigma\exp{(\gamma q_1)}\cos{\gamma q_2};\quad x_2=-\sigma\exp{(\gamma q_1)}\sin{\gamma q_2}; \nonumber\\
&&\tau_(q_1)+\tau_2(q_2)=\sigma^2\exp{(2\gamma q_1)}   \nonumber\\
&&V(x_1, x_2)=\frac{f(x_2/x_1)+g(x_1^2+x_2^2)}{x_1^2+x_2^2}, \label{SSS1}
\ea
coinciding with the case I of Eisenhart \cite{eisenhart}.

The second option with real $\sigma =-\delta=a/2$ gives exactly the case II of Eisenhart:
\ba
&&x_1=a\cosh{(\gamma q_1)}\cos{(\gamma q_2)};\quad x_2=-a\sinh{(\gamma q_1)}\sin{(\gamma q_2)};\nonumber\\
&&\tau_1(q_1)+\tau_2(q_2)=\frac{\gamma^2a^2}{2}[\cosh{(2\gamma q_1)}-\cos{(2\gamma q_2)}];\nonumber\\
&&V(x_1, x_2)=\frac{a^2[f(A+B)+g(B-A)]}{(A^2+x_2^2/a^2)^{1/2}};\,\, A\equiv \frac{x_1^2+x_2^2-a^2}{2a^2};
\,\, B\equiv (A^2+x_2^2/a^2)^{1/2}. \label{SSS2}
\ea

After the analysis above, it is clear that $R-$separation of variables for polynomial potential in (\ref{H}),
if possible at all, would belong to the option (\ref{alpha}). But explicit
substitution of (\ref{alpha}) into (\ref{Hz}) shows that this expression is not reducible to the form
(\ref{R}), i.e. the system (\ref{H}) is not amenable to separation of variables.

\section*{\normalsize\bf 5.\quad Conclusions.}
\vspace*{0.5cm}
\hspace*{3ex}
It is interesting to compare the non-Hermitian model (\ref{H}), (\ref{connection}), which
does not allow for separation of variables, with the same model, but without condition (\ref{connection}),
i.e. with the model
of \cite{srilanka}. It is clear that our model corresponds to merging of pairs of mutually complex conjugate
energy levels in \cite{srilanka}, since the restriction (\ref{connection}) just leads to vanishing imaginary
parts of energy eigenvalues. Thus, the condition (\ref{connection}) on coupling constant violates diagonalizability
of the Hamiltonian Eq.(3) in \cite{srilanka} and also turns complex energy levels to the real axis. The special remark
concerns the dimensionality of Jordan cells: $p_n=(n+1)$ coincides with the degeneracy of the corresponding
levels $E_n$ in the model (\ref{H}), for $g=0$ and $\omega_1=\omega_2,$ i.e. in the model of
isotropic real two-dimensional oscillator.
It would be interesting to investigate further the classical integrals of motion
and most importantly quantum symmetry operators
for the diagonalizable and non-diagonalizable cases, both being solvable,
and the interplay with separability of variables.

\section*{\normalsize\bf Acknowledgements}
The work was partially supported by INFN and University of Bologna (M.V.I. and D.N.N.), by Russian grants
RFFI 09-01-00145-a, RNP 2.1.1/1575 (M.V.I.) and by grant ATSU/09317 (D.N.N.).
\vspace{.2cm}

\section*{\normalsize\bf A1.\quad Symmetry operator.}
\vspace*{0.5cm}
\hspace*{3ex}
It follows from the shape invariance property, i.e. from intertwining relations (\ref{shape}),
that the operator $R=A^+A^-$ commutes with Hamiltonian $H.$
This fact allows to choose the wave functions of $H$ such that they are
simultaneously the eigenfunctions of $R.$
Let us solve the corresponding pair of equations:
 \ba
 (\partial_{z}^2-\frac{\lambda^2}{4}\bar{z}^2)\Psi=-\frac{\lambda^2}{4}r^2\Psi,\label{1}\\
 (-4\partial_{z}\partial_{\bar{z}}+\lambda^2z\bar{z}+g\bar{z}^2)\Psi=E\Psi,\label{2}
 \ea
where eigenvalues of $R$ are denoted as $-\lambda^2r^2/4$ with
arbitrary (complex) constant $r.$ The equation (\ref{1}) has two independent solutions:
 \ba
 \Psi^{(1)}(z, \bar z)=c^{(1)}(\bar{z})\exp(zf(\bar{z})), \qquad
 \Psi^{(2)}(z, \bar z)=c^{(2)}(\bar{z})\exp(-zf(\bar{z})),\label{3}
 \ea
where $c^{(1,2)}(\bar{z})$ are arbitrary functions, and
$f(\bar{z}) \equiv \lambda\sqrt{\bar{z}^2-r^2}/2.$ Substitution of (\ref{3})
into (\ref{2}) leads to first order homogeneous equations for
$c^{(1,2)}(\bar{z}):$
 \ba
 -4(c^{(1)}f)^{\prime}+(g\bar{z}^2-E)c^{(1)}(\bar{z})&=&0, \label{4}\\
 4(c^{(2)}f)^{\prime}+(g\bar{z}^2-E)c^{(2)}(\bar{z})&=&0. \label{5}
 \ea
They are solvable explicitly, in a general form:
 \ba
 c^{(1)}(\bar{z})=\frac{const}{f}\exp(\int\frac{g\bar{z}^2-E}{4f}d\bar{z}),\quad
 c^{(2)}(\bar{z})=\frac{const}{f}\exp(-\int\frac{g\bar{z}^2-E}{4f}d\bar{z}).
 \ea
It is easy to check that $\Psi^{(2)}(z, \bar z)$ has an exponentially decreasing
asymptotics at infinity. The apparent singularity of $c^{(2)}(\bar z)$ at zeros of $f(\bar z)$
can be compensated for $r=0$ just for the positive values of $E,$ thus justifying
the choice of positive $n$ in (\ref{spectrum}). For these values of
energy, the functions $\Psi^{(2)}(z, \bar z)$ coincide with wave functions
(\ref{formal}) in Section 2.

\end{document}